\documentclass[%
 reprint,
 amsmath,amssymb,
 aps,
]{revtex4-2}

\usepackage{graphicx}
\usepackage{dcolumn}
\usepackage{bm}

\begin{document}

\preprint{APS/123-QED}

\title{Response of CsI[Na] to Nuclear Recoils: \\
Impact on Coherent Elastic Neutrino-Nucleus Scattering (CE$\nu$NS)}

\author{J.I. Collar}
\email{collar@uchicago.edu}
\author{A.R.L. Kavner}
\author{C.M. Lewis}%
\affiliation{%
Enrico Fermi Institute, Kavli Institute for Cosmological Physics, and Department of Physics\\
University of Chicago, Chicago, Illinois 60637, USA
}%


\date{\today}

\begin{abstract}
A new measurement of the quenching factor for low-energy nuclear recoils in  CsI[Na] is presented. Past measurements are revisited, identifying and correcting several systematic effects. The resulting global data are well-described by a physics-based model for the generation of scintillation by ions in this material, in  agreement with phenomenological considerations. The uncertainty in the new model is reduced by a factor of four with respect to an energy-independent quenching factor initially adopted as a compromise by the COHERENT collaboration. A significantly improved agreement with Standard Model predictions for the first measurement of CE$\nu$NS is generated. We emphasize the critical impact of the quenching factor on the search for new physics via CE$\nu$NS experiments.
\end{abstract}

\maketitle


\section{\label{sec:level1}Introduction}

Low-energy nuclear recoils have been a subject of intense recent interest in experimental particle physics. Its onset can be traced back to the first direct search \cite{ahlenfta} for Weakly Interacting Massive Particles (WIMPs), dark matter candidates expected to interact via elastic scattering off nuclei in target detectors. The ensuing nuclear recoil (NR),  neutral or lightly ionized at the few keV energies of interest, rapidly dissipates its energy through a combination of secondary recoils (i.e., heat), and ionization. The first mechanism dominates for this NR energy range, leading to a markedly different path of energy quenching when compared to an electron recoil (ER) of similar energy, which will instead favor direct ionization. For particle detectors based on the generation of scintillation or free charge, measurable signals from NRs are consistently smaller than for ERs of the same energy. The ratio of their magnitudes, an energy-dependent quantity, is dubbed a ``quenching factor" for NRs.

Neutrinos with energies below few tens of MeV can scatter coherently from nuclei, also producing  few-keV NRs as the single observable, albeit with a dramatically enhanced probability vis-a-vis other  modes of interaction mediated by charged currents. This process,  referred to as Coherent Elastic Neutrino-Nucleus Scattering (CE$\nu$NS), was first proposed in 1974 \cite{freedman}. Due to the intrinsic difficulty in measuring low-energy nuclear recoils, it was not until 2017 that the first CE$\nu$NS detection was realized 
\cite{science,ournim,bjorn}.

Knowledge of the quenching factor (QF), typically acquired via detector calibrations involving neutron scattering, is essential for the interpretation of WIMP and CE$\nu$NS experimental data. In the first case, for as long as dark matter search results remain negative, the urgency of this knowledge may be temporarily de-stressed, as it only reflects on the exact region of parameter space excluded by the experiments, and on their relative sensitivities \cite{myprcnai}. For CE$\nu$NS studies, where an actual neutrino signal is registered, the importance of acquiring quality information about detector response cannot be overemphasized: as we will argue in Sec.\ VI, faulty information can lead to misidentifying apparent deviations from Standard Model CE$\nu$NS predictions as signatures of new physics, or conversely, to concealing such harbingers. 

In this work we revise the QF adopted in \cite{science} for CsI[Na] inorganic scintillator. An unphysical, energy-independent QF of 8.78\% was recommended there for the entire 5-30 keV energy range, the region of interest (ROI) for the NRs expected during this first observation of CE$\nu$NS. This tentative value was offered as a compromise, in face of the large disagreements between neutron calibration data available at the time. In Sec.\ II  we present a new measurement (``Chicago-3") of the QF for this material in this ROI, casting light on the situation. Sec.\ III  identifies a systematic effect impacting previous measurements (``Chicago-2" \cite{bjorn} and ``Duke" \cite{grayson}), correcting for it. In Sec.\ IV we re-analyze data from an earlier measurement (``Chicago-1"). Its original treatment \cite{ournim} used a definition of the QF that did not allow for a direct comparison to other measurements, or to contribute to the interpretation of CE$\nu$NS data from \cite{science}. These limitations are addressed in the present re-analysis. In Sec.\ V we show that the global dataset generated, comprising all six presently available CsI[Na] QF calibrations, is finally harmonious, within  experimental uncertainties. Most importantly, this data ensemble  follows a physics-based modified Birks model of scintillation by ions in CsI[Na], with best-fit parameters in good agreement with values derived from first principles. In Sec.\ VI we present an improved agreement between COHERENT data and Standard Model CE$\nu$NS predictions brought about by our new QF model. We conclude with a commentary on how this agreement and the considerably smaller uncertainty  derived from this work impacts searches for new physics within the COHERENT CsI[Na] dataset, providing a few specific examples. We finish by  illustrating the importance of best-effort detector calibration work, in the context of future CE$\nu$NS searches for neutrino magnetic moments using point-contact germanium detectors.

\section{New measurement of the  quenching factor for CsI[Na]}

A left panel in Fig.\ 1 shows the available neutron calibration data used in \cite{science} to suggest a QF value of 8.78\% for the CsI[Na] NR range 5-30 keV. A large uncertainty of 18.9\% was assigned, in view of the disagreements visible in the figure. In an effort to investigate the origin of these tensions, we performed a new calibration, using the same small 14.5 cm$^{3}$ CsI[Na] scintillator as before (Chicago-1, Chicago-2, Duke). This crystal was procured from the manufacturer \cite{proteus} of the 14.6 kg detector used in \cite{science} for CE$\nu$NS detection, with same sodium dopant concentration and growth method for both. A new ultra-bialkali (UBA) photomultiplier (PMT) of the same type used for Chicago-2 and Duke calibrations (Hamamatsu H11934-200) was employed. The Thermo MP320 D-D neutron generator, Bicron 501A backing detector, data-acquisition system, and triggering configuration were all in common to \cite{ournim,myprcnai}. The monochromatic 2.24 MeV (0.25 MeV FWHM) energy of the 2.5$\times10^{6}$ neutrons per second emitted from this source was measured in \cite{myprcnai} using a Cuttler-Shalev spectrometer. This energy is in good agreement with the $\sim$2.18 MeV expected from operation of a D-D generator at the 80 kV accelerating potential involved \cite{crc}. This calibration setup was utilized in \cite{myprcnai} to observe a monotonically decreasing QF with decreasing Na recoil energy in NaI[Tl], a trend at the time in tension with previous measurements. This behavior was later confirmed in four independent experiments \cite{nai1,nai2,nai3,nai4}.

\begin{widetext}
\begin{figure*}[!htbp]
\includegraphics[width=1 \linewidth]{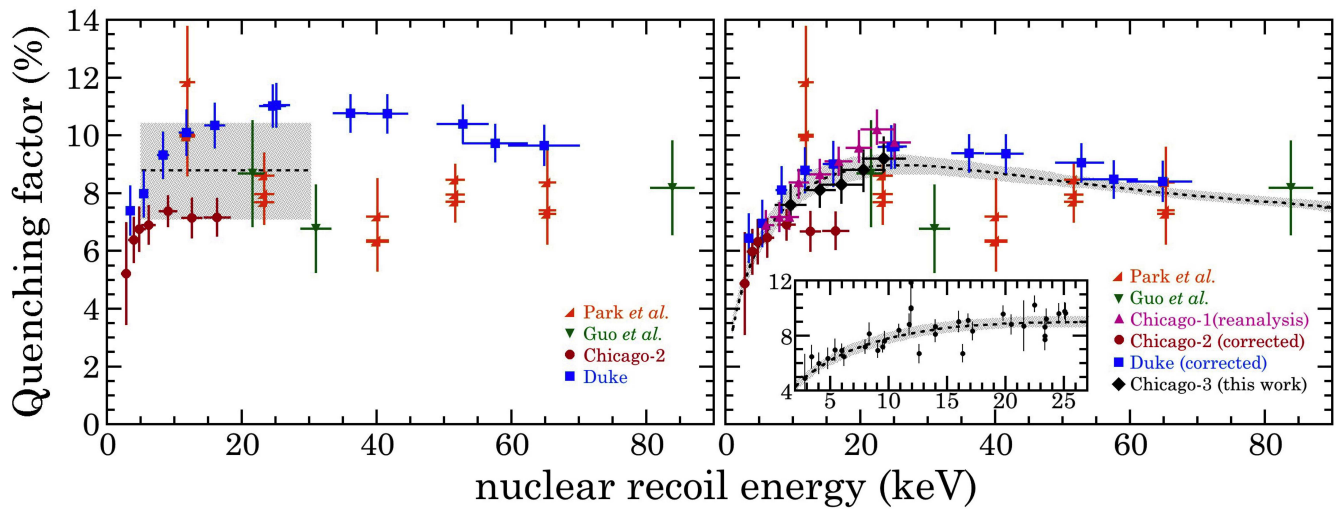}
\caption{\label{fig:wide}{\it Left:}  quenching factor suggested in \cite{science} for CsI[Na]. An unphysical energy-independent behavior was adopted to accommodate the large dispersion in calibration data available at the time \cite{bjorn,grayson,guo,park}, visible in the figure. The resulting  1-$\sigma$ uncertainty is shown as a grayed band. {\it Right:} present global data and physics-based QF model developed in Sec.\ V (dotted line). The inset expands the  CE$\nu$NS ROI for CsI[Na] at a  spallation source. Horizontal error bars are removed for clarity. }
\end{figure*}
\end{widetext}

\begin{figure}[!htbp]
\includegraphics[width=1 \linewidth]{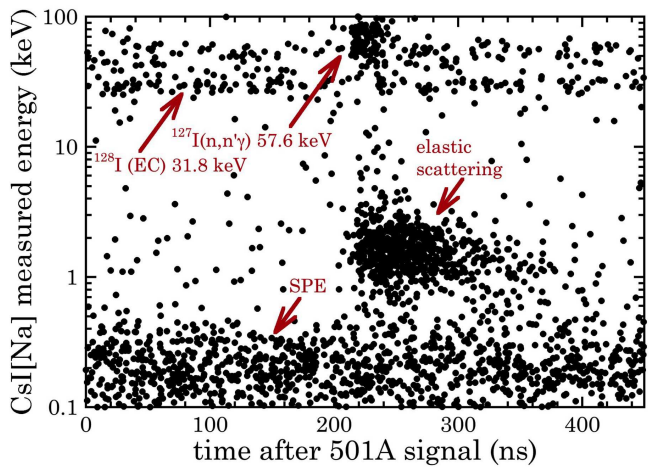}
\caption{\label{fig:epsart} Data quality in the new QF measurement for cesium and iodine recoils in CsI[Na], illustrated here for the  65$^{\circ}$ neutron scattering angle. Prompt coincidences between the CsI[Na] crystal under 2.24 MeV neutron irradiation and the Bicron 501A backing detector occur at $\sim$220 ns in the  horizontal time scale \cite{myprcnai}. The spilling of  elastic scattering events to later times is due to a $\sim$600 ns CsI[Na] scintillation decay constant \cite{ournim} and the limited light-yield (15.4 PE/keV) involved. Events from neutron capture, inelastic scattering, and afterglow comprised of up to a few PE, are also indicated. }
\end{figure}

The reader is referred to \cite{ournim,myprcnai} for a detailed description of this system. A few significant departures from the methodology described there were implemented. First, the energy reference in the definition of the QF was provided by the 59.54 keV gamma emission from a $^{241}$Am calibration source, as opposed to Compton scattering measurements over a broader range of ER energies. This was done to provide consistency with the energy scale in \cite{science,bjorn,grayson}, also based on $^{241}$Am exposure.  Second, the data analysis pipeline was revised to ensure an identical treatment in the extraction of PMT charge from single photoelectrons (SPE), $^{241}$Am signals, and NR signals. For the last two, charge was integrated over the 3 $\mu$s following the onset of CsI[Na] signals, as was done in \cite{science,bjorn,grayson}. Special attention was paid to the removal of low-amplitude PMT noise from digitizer traces. Third, the separation of neutron- and gamma-induced events in the  501A liquid scintillator cell was ameliorated through an integrated rise-time (IRT) analysis \cite{irt1,irt2}, resulting in an optimized removal of gamma backgrounds. Fig.\ 2 illustrates data quality following this IRT cut. Secondary improvements included a more detailed  geometry in MCNP-PoliMi simulations \cite{polimi}, and their accounting for inelastic  scattering with de-excitation gamma escape from CsI[Na]. Contributions from this process become non-negligible at the largest scattering angles considered. 

\begin{figure}[!htbp]
\includegraphics[width=.92 \linewidth]{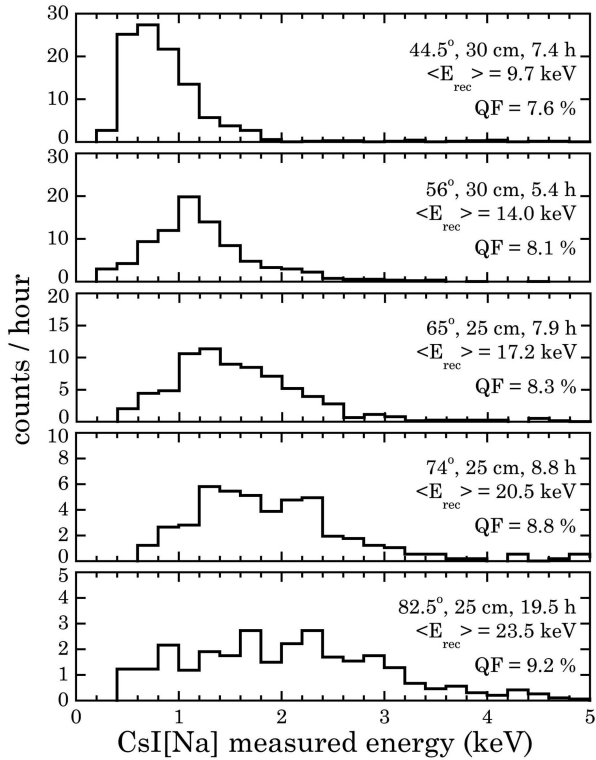}
\caption{\label{fig:epsart}  Measured energy for NRs from neutron scattering. Labels state scattering angle, distance between CsI[Na] and backing detector, exposure time to the D-D generator, simulated mean NR energy $<\!\!E_{rec}\!\!>$, and best-fit QF obtained from the analysis (see text). The decrease in event rate with larger angle is characteristic of forward-peaked elastic scatters.  }
\end{figure}

\begin{figure}[!htbp]
\includegraphics[width=1 \linewidth]{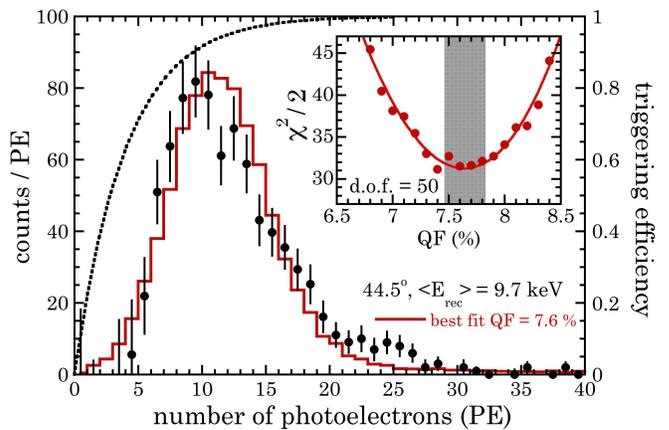}
\caption{\label{fig:epsart} Comparison of NR light-yield obtained for the smallest scattering angle tested, and its best-fit simulated prediction. A dotted line shows the triggering efficiency of the calibration setup, derived as in \cite{myprcnai,ournim}. A correction based on this efficiency is applied to the datapoints shown, prior to comparison to simulations. A greyed band in the inset shows the 1-$\sigma$ uncertainty in the best-fit QF value (see text). }
\end{figure}

Random coincidences between CsI[Na] and backing detector are removed during data processing by subtraction of the energy spectrum for events in the 40-220 ns time range of Fig.\ 2, from that for those in the 220-400 ns interval, the latter being dominated by true coincidences. The residual CsI[Na] energy spectra thus obtained are shown in Fig.\ 3, where the sought elastic scattering NR signals are cleanly isolated. Simple inspection of the measured energy for NRs (derived from a comparison to reference $^{241}$Am gammas) and the simulated mean NR energy, points to QF values of order 8\%. The actual extraction of a best-fit QF, for each of five scattering angles tested, is accomplished by translating measured energies to an equivalent light-yield, expressed in number of photoelectrons (PE) at the CsI[Na] PMT. In these units, we can account for Poisson smearing of simulated energy depositions involving a  small expected number of PE. 

Fig.\ 4 shows a comparison of experimental NR light-yield (datapoints) and its best-fit simulated prediction (histogram). The fit is obtained by scanning the 5\%-15\% QF range, using a standard log-likelihood approach. An estimate of the uncertainty in the QF values obtained is computed from the sum of the 1-$\sigma$ log-likelihood error (Fig.\ 4, inset), and the dispersion in the $^{241}$Am light-yield (15.4 $\pm$ 1.2 PE/keV) obtained across six different determinations of this energy reference, performed before and after each scattering angle run. This  uncertainty appears as vertical error bars for the Chicago-3 datapoints in the right panel of Fig.\ 1. Horizontal error bars there provide a measure of the recoil energies probed by each run. They result from the angle subtended by the backing detector from the position of the CsI[Na] crystal, and the kinematic relationship between neutron scattering angle and NR energy \cite{ej301}.

\section{Impact of PMT saturation\\ on previous measurements}

A comparison of results from the new calibration described above (Chicago-3, Fig.\ 1) with uncorrected previous Chicago-2  and Duke  values (Fig.\ 1, left panel) reveals considerable remaining disagreement. Prompted by this, we examined the differences between these measurements. Chicago-2 \cite{bjorn} and Duke \cite{grayson} calibrations were performed within days of each other, using the same PMT and its mounted CsI[Na] crystal, and same 3.85 MeV (0.37 MeV FWHM) monochromatic neutron source \cite{tunldd}, but employing different backing detectors, data-acquisition system, and analysis. Noticing that a high-voltage bias just 50 V below maximum ratings had been applied to the CsI[Na] PMT in \cite{grayson}, we considered the possibility of PMT saturation at the corresponding high gain. The large current induced at the last PMT dynode stages under such extreme conditions can increase the local space charge density, resulting in a current output limitation (PMT ``saturation" or ``non-linearity") even for relatively small instantaneous light inputs \cite{handbook,report}. This scenario might have led to a stunted current output for the $^{241}$Am energy reference (generating close to a 1,000 PE input to the PMT), while not affecting the much smaller (few to few tens of time-separated SPE) signals from neutron-induced NRs. The net effect of this, when gone unnoticed, would be the extraction of artificially large QF values. 

\begin{figure}[!htbp]
\includegraphics[width=1 \linewidth]{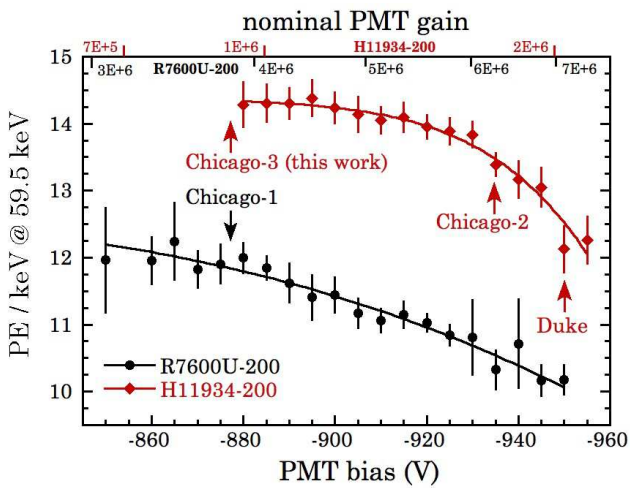}
\caption{\label{fig:epsart} Tests of UBA PMT saturation effect under $^{241}$Am irradiation of a CsI[Na] crystal. The operating bias used in four QF calibrations described in the text is indicated by labelled arrows. The top axis shows PMT gain, derived from manufacturer specifications. The low-bias range of these measurements was defined by the ability to clearly separate unamplified SPEs from PMT noise. A logistic function is used to fit the data (solid lines). Error bars merge the uncertainties from fits to SPE and  $^{241}$Am charge distributions (see text). }
\end{figure}

 Due to the relatively long fast-decay constant of CsI[Na] ($\sim$550 ns, \cite{ournim,decay}), able to protract the light input to the PMT, this effect might be modest. Acknowledging this, we tested our hypothesis, using the CsI[Na] crystal common to all calibrations listed in Fig.\ 5, and two UBA PMTs. The first (Hamamatsu R7600U-200) was the same unit as in the Chicago-1 calibration discussed in the next section. The second (Hamamatsu H11934-200) was a different unit of the same model used for Chicago-2 and Duke calibrations, as the original PMT was unavailable to us. The present unit was used for the Chicago-3 calibration described above. Measurements of  $^{241}$Am current output were performed at a variety of PMT biases, comparing them with the charge distribution for SPEs in each run. Their ratio provides the $^{241}$Am light-yield in PE/keV, as a function of bias. 
 
 Fig.\ 5 shows the results from these tests, confirming the presence of a measurable saturation effect. A correction factor for the QF values obtained in the different calibrations can be derived from the ratio of light-yield at operating PMT bias, and the asymptotic light-yield at low bias, where saturation is absent, as is expected for low-amplitude NR signals. This correction is maximal for Duke runs (0.87), smaller for Chicago-1 and Chicago-2 (0.92 and 0.94, respectively), and not applicable to Chicago-3, where the bias applied was sufficiently low to avoid this systematic effect. While some variability in saturation behavior can be expected between units of the same PMT model \cite{report}, we notice that $^{241}$Am light-yields obtained for the H11934-200 originally used in Chicago-2 and Duke calibrations  (13.4 PE/keV in \cite{bjorn}, 12.0 PE/keV in \cite{grayson,pekev}) are essentially identical to those in Fig.\ 5 for our unit. This makes us confident in the magnitude of the corrections implemented. 
 
 Revised QF values for Duke and Chicago-2 calibrations can be observed in the right panel of Fig.\ 1, now in much better agreement between themselves and with Chicago-3. The saturation effect described in this section should be absent from the CE$\nu$NS data in \cite{science}, as the Hamamatsu R877-100 PMT employed there was intentionally operated at a modest gain of 7$\times10^{5}$, relying on external amplification for SPE counting \cite{bjorn}.

\section{Re-analysis of Chicago-1 data}

The original Chicago-1 calibration returned monotonically increasing QF values in the range 4-7 \%, for increasing NR energy in the interval 6-27 keV \cite{ournim}. The ER energy reference was provided by a dedicated measurement of Compton scattering from a collimated $^{133}$Ba gamma source, using a germanium diode as backing detector \cite{myprcnai,ournim}. This allowed a direct comparison of the magnitude of NR and ER measured signals corresponding to the same energy, instead of using a fixed 59.5 keV ER reference energy, as above. This alternative definition of the QF is of interest when an intrinsic low-energy ER reference exists, as is the case for internal 3.2 keV emissions from $^{40}$K in NaI[Tl] dark matter detectors \cite{myprcnai}. 

The specific definition of the QF is a matter of convention. For instance, for dark matter liquid xenon detectors the QF energy reference is traditionally, but not always, 122 keV ERs from $^{57}$Co \cite{leff}. However, internal consistency in the definition of an energy scale is necessary, if CsI[Na] QF calibrations are to be used in the interpretation of CsI[Na] CE$\nu$NS data. For CsI[Na], NRs and ERs display an opposite trend in the dependence of their scintillation decay constants on energy, noticeable in Fig.\ 6 of \cite{ournim}. This was of particular concern when contemplating the adoption of Chicago-1 QF values for the interpretation of COHERENT CE$\nu$NS data: if signals from NRs and ERs of the same energy are compared, this opposite trend can exaggerate the decrease of the QF with lowering energy, when using a finite (3 $\mu$s) scintillation integration window. In view of this, as a precaution, Chicago-1 data were excluded from consideration in \cite{science}.

Noticing the large remaining disagreement between Chicago-1 QF values as shown in \cite{ournim}, and the now corrected Duke, Chicago-2, and new Chicago-3 values, we attempted to estimate the effect of ER and NR decay constants just mentioned, concluding that it has an insufficient impact (few \%) to alleviate this tension. Prompted by this, we applied the new analysis tools developed in Sec.\ II, to re-analyze Chicago-1 data, anew. In the absence of data from $^{241}$Am exposure, we used the largest Compton scattering angle (36$^{\circ}$) measured in \cite{ournim}, which produced a broad (18 keV FWHM) distribution of CsI[Na] ER energies centered around 41.8 keV. Existing studies of gamma and electron scintillation non-proportionality \cite{nonprop1,nonprop2,nonprop3} indicate that differences in light-yield  per unit energy between 41.8 keV and 59.5 keV are negligible for this material. This sanctions our alternative energy reference as a good facsimile of $^{241}$Am.

The light-yield distributions for NRs obtained from this full re-analysis are very similar to those in Fig.\ 4 of \cite{ournim}, shifted to a higher number of PE by just 10-20 \%. This change is again insufficient to resolve the tension with other QF calibrations. However, the new light-yield for ERs was found to be considerably smaller at 13.7 PE/keV than the $\sim$17 PE/keV in \cite{ournim}. This change was not limited to the 36$^{\circ}$ Compton dataset, indicating that a mistake or systematic in computing Compton ER light-yields affected the original Chicago-1 QF determination in \cite{ournim}. Perhaps not surprisingly, we find that the re-analyzed Chicago-1 QF values are now in excellent agreement with all other calibrations discussed (Fig.\ 1, right panel). The small correction for PMT saturation derived from Fig.\ 5 for Chicago-1 is included in our re-analysis, while keeping in mind that photocathode aging over a decade, and remounting of crystal and PMT can lead to small divergences in the light-yield measured. 

\section{A physics-based model of CsI[Na]\\
scintillation by slow ions}

The right panel in Fig.\ 1, containing all revised and new QF data, reveals what is seemingly a good harmony between all contributions, limited only by experimental uncertainties. This begs for further interpretation. 

The subject of physics-based  models for the generation of scintillation by low-energy NRs in a variety of targets, including CsI[Na], has been most recently reviewed in \cite{tretyak}. The model favored there is a well-accepted semi-empirical approach by Birks \cite{birks}, containing a single free parameter $kB$. In the low-energy approximation proposed in \cite{tretyak}, the fractional quenching factor is simply given by QF $=(kB\cdot(dE/dr)_{i})^{-1}$, where $(dE/dr)_{i}$ is the energy-dependent total stopping power for ions in scintillator. While such a model was seen in \cite{tretyak} to satisfactorily explain the scarce CsI[Na] QF data available in 2010 \cite{park}, it predicts a monotonically increasing QF with decreasing energy, at clear variance with present data. 

An important component of the microphysics governing scintillation generation by slow ions is unfortunately absent from the treatment in \cite{tretyak}.  As described in \cite{ahlen1,ahlen2,ziegler}, basic two-body kinematic considerations predict a cutoff to the production of scintillation whenever the maximum possible energy transfer to an electron by a slow-moving NR falls below the minimum excitation energy (band gap, $E_{g}$) specific to the material. In a Fermi-gas model for atomic electrons, this condition for scintillation production can be written  as $2m_{e}v(v+v_{F})>E_{g}$ \cite{ahlen1}, where $v$ is the ion velocity, $m_{e}$ is the electron mass, and the Fermi velocity of atomic electrons is given by $v_{F}=(\hbar/m_{e})(3\pi^{2}\rho)^{1/3}$ \cite{ziegler}, with $\rho$ as the electron density of the medium. This cutoff is not expected to be abrupt, due to high-velocity tails in the electron momenta distributions \cite{ahlen1,ahlen2}. To account for this smooth transition, an adiabatic factor of the form 1-exp(-$E_{rec}/E_{0}$) can be introduced, where $E_{rec}$ is the NR energy, and $E_{0}$ the limiting ion kinetic energy that fails to meet the inequality above. This factor was used in \cite{ahlen1} to successfully modify the QF model for proton recoils in organic plastic scintillator, accounting for experimental observations. 

 Following the discussion above, we adopt a simple model of scintillation by slow ions in CsI[Na], described by the product of the low-energy approximation to Birks derived in \cite{tretyak}, and an adiabatic factor. This involves just two free parameters, $kB$ and $E_{0}$. As in \cite{tretyak}, we extract the total stopping power for ions $(dE/dr)_{i}$ from SRIM-2013 \cite{srim}. We fit this model to the global data in the right panel of Fig.\ 1, using a popular Markov Chain Monte Carlo (MCMC) ensemble sampler \cite{mcmc1,mcmc2}. For simplicity, horizontal error bars are neglected, as they do not represent an uncertainty, but instead the approximate  span of NR energies explored by the calibrations. The sampler rapidly converges  to best-fit values of $kB = 3.311\pm0.075\times10^{-3}$ g\! /\! MeV \!cm$^{2}$  and $E_{0} = 12.97 \pm 0.61$ keV, where the errors  define the 1-$\sigma$ uncertainties in the model (grayed band in Fig.\ 1). The value of $kB$ is unremarkable, being comparable to those found for a number of inorganic scintillators in \cite{tretyak}. 
 
 Much more interesting is the returned best-fit value for $E_{0}$: using the expressions above, we independently estimate that $E_{0}$ for Cs and I recoils in CsI[Na] should fall in the interval 11-14 keV, depending on which value of $E_{g}$ is adopted (5.5-6.2 eV \cite{bandgap1,bandgap2}). Not only the global QF data for CsI[Na] are presently in agreement, but they are also accurately described by an energy-dependent physical model, one that can be derived from first principles.

\section{Conclusions}

The drastic reduction in QF uncertainty expressed by the contrast between left and right panels in Fig.\ 1 is expected to have a large impact on phenomenological investigations exploiting the CsI[Na] COHERENT dataset. The original 18.9\% QF uncertainty dominated the total uncertainty budget for Standard Model (SM) CE$\nu$NS predictions: at 5.1\% in the CE$\nu$NS ROI, its effect is now comparable to the choice of theoretical nuclear form factor in signal calculations \cite{science,danny}.  What is more, the generally smaller new QF generates a SM CE$\nu$NS predicted rate ($\sim$138$\pm$19 events) much closer to the observed signal (134$\pm$22 events \cite{science}) than the original expectation (173$\pm$48 events \cite{science}, Fig.\ 6). The synergistic combination of a shift towards better agreement between SM and experiment, and the factor-of-two drop in overall uncertainty (after accounting for all other sources listed in \cite{science}), should provide a considerably increased sensitivity to new physics. The new QF model  will maximize the usefulness of an upcoming final CsI[Na] COHERENT dataset,  twice the original exposure.

\begin{figure}[!htbp]
\includegraphics[width=.93 \linewidth]{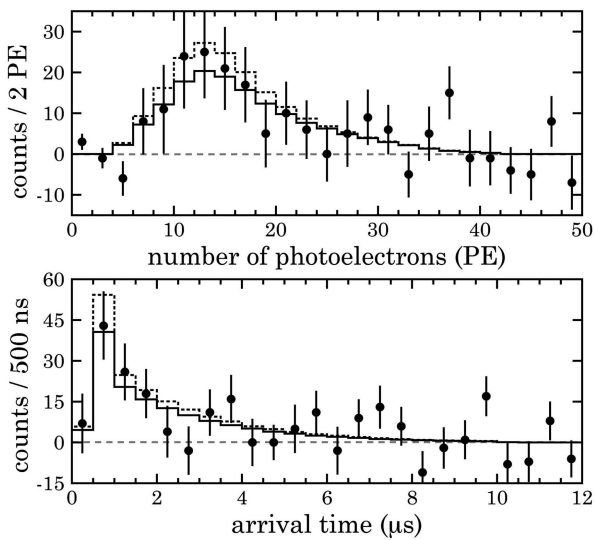}
\caption{\label{fig:epsart} COHERENT CsI[Na] CE$\nu$NS signal (datapoints, \cite{science}) projected on energy and time after SNS proton spill, compared to previous SM predictions (dotted histogram) and a revision using the new QF model presented in this work (solid histogram). An improved agreement with the SM is obtained, with much smaller uncertainty in signal prediction (see text).}
\end{figure}

Studies on different aspects of neutrino and nuclear physics extracted from \cite{science} are too numerous to attempt an exhaustive assessment of the expected impact from this work. We offer just a few examples. For instance, the so-called LMA-D alternative to the more standard MSW picture of solar neutrino oscillations has been disfavored in \cite{lma,ian} at the  inconclusive $\sim$3-$\sigma$ level, based on bounds on neutrino non-standard interactions (NSI) derived from \cite{science}. The exclusion of LMA-D should be definitive now, through the improved NSI limits that are to be expected, a subject of interest in its own right. Similarly, a moderate tension of COHERENT CsI[Na] data with an effective weak mixing angle, described in a context of possible new physics \cite{huang1}, may be enhanced by a smaller QF value \cite{huang2}. To cite another example, a recent search for dark matter signatures in the CsI[Na] CE$\nu$NS dataset finds a mild excess at the 3.3-$\sigma$ statistical level, not explainable by  backgrounds \cite{strigari}. Interestingly, we find that the new QF model presented here does little to dispel this possibility, as it has a modest effect on the highest-energy CE$\nu$NS NRs constituting a background for that search. Finally, the now smaller separation between SM prediction and experimental CE$\nu$NS rate is better able to accommodate a $\sim$10\% signal rate contribution from incoherent neutrino scattering \cite{incoherent}.

The utility of CE$\nu$NS as a new tool in neutrino physics  only goes as far as our ability to understand the response of detector technologies to low-energy ions. We conclude by further illustrating this point with Fig.\ 7, where the impact of a finite neutrino magnetic moment on CE$\nu$NS spectra, and two alike models of germanium QF, are seen to compete for the interpretation of future data. In preparation for the use of p-type point contact (PPC) germanium detectors \cite{ppc} in CE$\nu$NS studies, we have explored the sub-keV germanium QF, combining a new exposure to $^{88}$Y/Be and $^{124}$Sb/Be monochromatic neutron sources \cite{ybeprl,bjornqf,alvaro}, with a repetition of a reactor measurement yielding 254 eV NRs \cite{jones,ohio}, and an improved calibration using 24 keV neutrons from a filtered beam \cite{ppc,ksu}. The picture provided by this body of new information will be the subject of a future publication.

\begin{figure}[!htbp]
\includegraphics[width=.98 \linewidth]{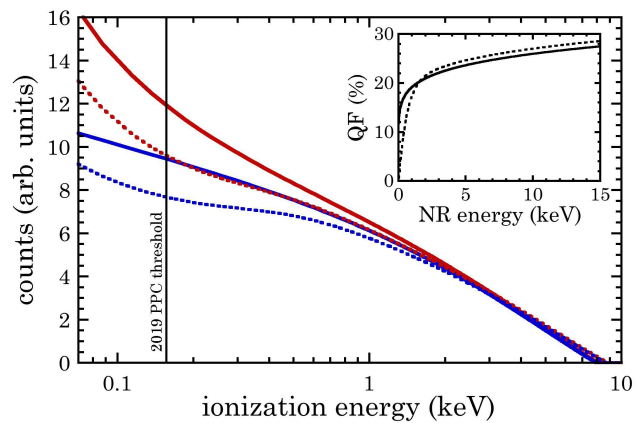}
\caption{\label{fig:epsart} CE$\nu$NS signals on germanium at a spallation source. A vertical line shows the best contemporary threshold for 1 kg PPCs \cite{cdex}. Red traces correspond to $\mu_{\nu_{\mu}}=5\times10^{-10}\mu_{Bohr}$, beyond the present limit of $\mu_{\nu_{\mu}}<6.8\times10^{-10}\mu_{Bohr}$ \cite{lsnd}. Blue lines assume zero neutrino magnetic moment. Solid lines correspond to a Lindhard QF model with free parameter $k=0.18$  (inset, \cite{bjornqf}). Dotted lines add an adiabatic factor (Sec.\ V, $E_{0}=0.5$ keV)  to a $k=0.19$ Lindhard model. The degeneracy of possible interpretations for hypothetical data following the two overlapping lines is evident: new physics could be erroneously claimed with high confidence, or missed while blatantly present, depending on the QF model settled upon. }
\end{figure}

\begin{acknowledgments}
We are indebted to L.-W. Chen, Y. Efremenko, X.-R. Huang, A. Konovalov, and B. Scholz for useful exchanges. This work was supported by NSF awards PHY-1806722, PHY-1812702, and PHY-1506357. It was also partially supported by the Kavli Institute for Cosmological Physics at the University of Chicago through NSF Award 1125897, and an endowment from the Kavli Foundation and its founder Fred Kavli. 
\end{acknowledgments}

\bibliography{apssamp}

\providecommand{\noopsort}[1]{}\providecommand{\singleletter}[1]{#1}%
\begin{thebibliography}{54}%
\makeatletter
\providecommand \@ifxundefined [1]{%
 \@ifx{#1\undefined}
}%
\providecommand \@ifnum [1]{%
 \ifnum #1\expandafter \@firstoftwo
 \else \expandafter \@secondoftwo
 \fi
}%
\providecommand \@ifx [1]{%
 \ifx #1\expandafter \@firstoftwo
 \else \expandafter \@secondoftwo
 \fi
}%
\providecommand \natexlab [1]{#1}%
\providecommand \enquote  [1]{``#1''}%
\providecommand \bibnamefont  [1]{#1}%
\providecommand \bibfnamefont [1]{#1}%
\providecommand \citenamefont [1]{#1}%
\providecommand \href@noop [0]{\@secondoftwo}%
\providecommand \href [0]{\begingroup \@sanitize@url \@href}%
\providecommand \@href[1]{\@@startlink{#1}\@@href}%
\providecommand \@@href[1]{\endgroup#1\@@endlink}%
\providecommand \@sanitize@url [0]{\catcode `\\12\catcode `\$12\catcode
  `\&12\catcode `\#12\catcode `\^12\catcode `\_12\catcode `\%12\relax}%
\providecommand \@@startlink[1]{}%
\providecommand \@@endlink[0]{}%
\providecommand \url  [0]{\begingroup\@sanitize@url \@url }%
\providecommand \@url [1]{\endgroup\@href {#1}{\urlprefix }}%
\providecommand \urlprefix  [0]{URL }%
\providecommand \Eprint [0]{\href }%
\providecommand \doibase [0]{https://doi.org/}%
\providecommand \selectlanguage [0]{\@gobble}%
\providecommand \bibinfo  [0]{\@secondoftwo}%
\providecommand \bibfield  [0]{\@secondoftwo}%
\providecommand \translation [1]{[#1]}%
\providecommand \BibitemOpen [0]{}%
\providecommand \bibitemStop [0]{}%
\providecommand \bibitemNoStop [0]{.\EOS\space}%
\providecommand \EOS [0]{\spacefactor3000\relax}%
\providecommand \BibitemShut  [1]{\csname bibitem#1\endcsname}%
\let\auto@bib@innerbib\@empty
\bibitem [{\citenamefont {Ahlen}\ \emph {et~al.}(1987)\citenamefont {Ahlen},
  \citenamefont {Avignone}, \citenamefont {Brodzinski}, \citenamefont
  {Drukier}, \citenamefont {Gelmini},\ and\ \citenamefont
  {Spergel}}]{ahlenfta}%
  \BibitemOpen
  \bibfield  {author} {\bibinfo {author} {\bibfnamefont {S.~P.}\ \bibnamefont
  {Ahlen}}, \bibinfo {author} {\bibfnamefont {F.~T.}\ \bibnamefont {Avignone}},
  \bibinfo {author} {\bibfnamefont {R.~L.}\ \bibnamefont {Brodzinski}},
  \bibinfo {author} {\bibfnamefont {A.~K.}\ \bibnamefont {Drukier}}, \bibinfo
  {author} {\bibfnamefont {G.}~\bibnamefont {Gelmini}},\ and\ \bibinfo {author}
  {\bibfnamefont {D.~N.}\ \bibnamefont {Spergel}},\ }\href
  {https://doi.org/https://doi.org/10.1016/0370-2693(87)91581-4} {\bibfield
  {journal} {\bibinfo  {journal} {Phys.\ Lett.\ B}\ }\textbf {\bibinfo {volume}
  {195}},\ \bibinfo {pages} {603 } (\bibinfo {year} {1987})}\BibitemShut
  {NoStop}%
\bibitem [{\citenamefont {Freedman}(1974)}]{freedman}%
  \BibitemOpen
  \bibfield  {author} {\bibinfo {author} {\bibfnamefont {D.~Z.}\ \bibnamefont
  {Freedman}},\ }\href {https://doi.org/10.1103/PhysRevD.9.1389} {\bibfield
  {journal} {\bibinfo  {journal} {Phys. Rev. D}\ }\textbf {\bibinfo {volume}
  {9}},\ \bibinfo {pages} {1389} (\bibinfo {year} {1974})}\BibitemShut
  {NoStop}%
\bibitem [{\citenamefont {Akimov}\ \emph {et~al.}(2017)\citenamefont {Akimov}
  \emph {et~al.}}]{science}%
  \BibitemOpen
  \bibfield  {author} {\bibinfo {author} {\bibfnamefont {D.}~\bibnamefont
  {Akimov}} \emph {et~al.},\ }\href {https://doi.org/10.1126/science.aao0990}
  {\bibfield  {journal} {\bibinfo  {journal} {Science}\ }\textbf {\bibinfo
  {volume} {357}},\ \bibinfo {pages} {1123} (\bibinfo {year}
  {2017})}\BibitemShut {NoStop}%
\bibitem [{\citenamefont {Collar}\ \emph {et~al.}(2015)\citenamefont {Collar},
  \citenamefont {Fields}, \citenamefont {Hai}, \citenamefont {Hossbach},
  \citenamefont {Orrell}, \citenamefont {Overman}, \citenamefont {Perumpilly},\
  and\ \citenamefont {Scholz}}]{ournim}%
  \BibitemOpen
  \bibfield  {author} {\bibinfo {author} {\bibfnamefont {J.~I.}\ \bibnamefont
  {Collar}}, \bibinfo {author} {\bibfnamefont {N.~E.}\ \bibnamefont {Fields}},
  \bibinfo {author} {\bibfnamefont {M.}~\bibnamefont {Hai}}, \bibinfo {author}
  {\bibfnamefont {T.~W.}\ \bibnamefont {Hossbach}}, \bibinfo {author}
  {\bibfnamefont {J.~L.}\ \bibnamefont {Orrell}}, \bibinfo {author}
  {\bibfnamefont {C.~T.}\ \bibnamefont {Overman}}, \bibinfo {author}
  {\bibfnamefont {G.}~\bibnamefont {Perumpilly}},\ and\ \bibinfo {author}
  {\bibfnamefont {B.}~\bibnamefont {Scholz}},\ }\href
  {https://doi.org/10.1016/j.nima.2014.11.037} {\bibfield  {journal} {\bibinfo
  {journal} {Nucl.\ Instr.\ Meth.\ A}\ }\textbf {\bibinfo {volume} {773}},\
  \bibinfo {pages} {56 } (\bibinfo {year} {2015})}\BibitemShut {NoStop}%
\bibitem [{\citenamefont {Scholz}(2017)}]{bjorn}%
  \BibitemOpen
  \bibfield  {author} {\bibinfo {author} {\bibfnamefont {B.}~\bibnamefont
  {Scholz}},\ }\href@noop {} {Ph.D. thesis},\ \bibinfo  {school} {University of
  Chicago} (\bibinfo {year} {2017}),\ \Eprint
  {https://arxiv.org/abs/1904.01155} {arXiv:1904.01155} \BibitemShut {NoStop}%
\bibitem [{\citenamefont {Collar}(2013{\natexlab{a}})}]{myprcnai}%
  \BibitemOpen
  \bibfield  {author} {\bibinfo {author} {\bibfnamefont {J.~I.}\ \bibnamefont
  {Collar}},\ }\href {https://doi.org/10.1103/PhysRevC.88.035806} {\bibfield
  {journal} {\bibinfo  {journal} {Phys. Rev. C}\ }\textbf {\bibinfo {volume}
  {88}},\ \bibinfo {pages} {035806} (\bibinfo {year}
  {2013}{\natexlab{a}})}\BibitemShut {NoStop}%
\bibitem [{\citenamefont {Rich}(2017)}]{grayson}%
  \BibitemOpen
  \bibfield  {author} {\bibinfo {author} {\bibfnamefont {G.}~\bibnamefont
  {Rich}},\ }\href@noop {} {Ph.D. thesis},\ \bibinfo  {school} {University of
  North Carolina} (\bibinfo {year} {2017})\BibitemShut {NoStop}%
\bibitem [{pro()}]{proteus}%
  \BibitemOpen
  \href@noop {} {}\bibinfo {note} {Proteus Inc.\, Chagrin Falls, OH 44022, USA.
  Crystals manufactured by Amcrys-H, Ukraine.}\BibitemShut {Stop}%
\bibitem [{\citenamefont {Csikai}(1987)}]{crc}%
  \BibitemOpen
  \bibfield  {author} {\bibinfo {author} {\bibfnamefont {G.}~\bibnamefont
  {Csikai}},\ }\href@noop {} {\emph {\bibinfo {title} {\protect{CRC Handbook of
  Fast Neutron Generators}}}}\ (\bibinfo  {publisher} {\protect{CRC} Press},\
  \bibinfo {year} {1987})\BibitemShut {NoStop}%
\bibitem [{\citenamefont {{Rich}}\ \emph {et~al.}(2014)\citenamefont {{Rich}},
  \citenamefont {{Barbeau}}, \citenamefont {{Howell}},\ and\ \citenamefont
  {{Karwowski}}}]{nai1}%
  \BibitemOpen
  \bibfield  {author} {\bibinfo {author} {\bibfnamefont {G.}~\bibnamefont
  {{Rich}}}, \bibinfo {author} {\bibfnamefont {P.}~\bibnamefont {{Barbeau}}},
  \bibinfo {author} {\bibfnamefont {C.}~\bibnamefont {{Howell}}},\ and\
  \bibinfo {author} {\bibfnamefont {H.}~\bibnamefont {{Karwowski}}},\ }in\
  \href {https://doi.org/10.1103/BAPS.2014.APRIL.Y13.8} {\emph {\bibinfo
  {booktitle} {APS Meeting Abstracts}}}\ (\bibinfo {year} {2014})\ p.\ \bibinfo
  {pages} {Y13.008}\BibitemShut {NoStop}%
\bibitem [{\citenamefont {Xu}\ \emph {et~al.}(2015)\citenamefont {Xu} \emph
  {et~al.}}]{nai2}%
  \BibitemOpen
  \bibfield  {author} {\bibinfo {author} {\bibfnamefont {J.}~\bibnamefont {Xu}}
  \emph {et~al.},\ }\href {https://doi.org/10.1103/PhysRevC.92.015807}
  {\bibfield  {journal} {\bibinfo  {journal} {Phys. Rev. C}\ }\textbf {\bibinfo
  {volume} {92}},\ \bibinfo {pages} {015807} (\bibinfo {year}
  {2015})}\BibitemShut {NoStop}%
\bibitem [{\citenamefont {Stiegler}\ \emph {et~al.}(2017)\citenamefont
  {Stiegler}, \citenamefont {Sofka}, \citenamefont {Webb},\ and\ \citenamefont
  {White}}]{nai3}%
  \BibitemOpen
  \bibfield  {author} {\bibinfo {author} {\bibfnamefont {T.}~\bibnamefont
  {Stiegler}}, \bibinfo {author} {\bibfnamefont {C.}~\bibnamefont {Sofka}},
  \bibinfo {author} {\bibfnamefont {R.~C.}\ \bibnamefont {Webb}},\ and\
  \bibinfo {author} {\bibfnamefont {J.~T.}\ \bibnamefont {White}},\ }\href@noop
  {} {\  (\bibinfo {year} {2017})},\ \Eprint {https://arxiv.org/abs/1706.07494}
  {arXiv:1706.07494} \BibitemShut {NoStop}%
\bibitem [{\citenamefont {Joo}\ \emph {et~al.}(2019)\citenamefont {Joo},
  \citenamefont {Park}, \citenamefont {Kim}, \citenamefont {Lee}, \citenamefont
  {Kim}, \citenamefont {Kim}, \citenamefont {Lee},\ and\ \citenamefont
  {Kim}}]{nai4}%
  \BibitemOpen
  \bibfield  {author} {\bibinfo {author} {\bibfnamefont {H.~W.}\ \bibnamefont
  {Joo}}, \bibinfo {author} {\bibfnamefont {H.~S.}\ \bibnamefont {Park}},
  \bibinfo {author} {\bibfnamefont {J.~H.}\ \bibnamefont {Kim}}, \bibinfo
  {author} {\bibfnamefont {J.~Y.}\ \bibnamefont {Lee}}, \bibinfo {author}
  {\bibfnamefont {S.~K.}\ \bibnamefont {Kim}}, \bibinfo {author} {\bibfnamefont
  {Y.~D.}\ \bibnamefont {Kim}}, \bibinfo {author} {\bibfnamefont {H.~S.}\
  \bibnamefont {Lee}},\ and\ \bibinfo {author} {\bibfnamefont {S.~H.}\
  \bibnamefont {Kim}},\ }\href
  {https://doi.org/https://doi.org/10.1016/j.astropartphys.2019.01.001}
  {\bibfield  {journal} {\bibinfo  {journal} {Astropart. Phys.}\ }\textbf
  {\bibinfo {volume} {108}},\ \bibinfo {pages} {50 } (\bibinfo {year}
  {2019})}\BibitemShut {NoStop}%
\bibitem [{\citenamefont {Guo}\ \emph {et~al.}(2016)\citenamefont {Guo} \emph
  {et~al.}}]{guo}%
  \BibitemOpen
  \bibfield  {author} {\bibinfo {author} {\bibfnamefont {C.}~\bibnamefont
  {Guo}} \emph {et~al.},\ }\href
  {https://doi.org/https://doi.org/10.1016/j.nima.2016.02.037} {\bibfield
  {journal} {\bibinfo  {journal} {Nucl. Instr. Meth. A}\ }\textbf {\bibinfo
  {volume} {818}},\ \bibinfo {pages} {38 } (\bibinfo {year}
  {2016})}\BibitemShut {NoStop}%
\bibitem [{\citenamefont {Park}\ \emph {et~al.}(2002)\citenamefont {Park} \emph
  {et~al.}}]{park}%
  \BibitemOpen
  \bibfield  {author} {\bibinfo {author} {\bibfnamefont {H.}~\bibnamefont
  {Park}} \emph {et~al.},\ }\href
  {https://doi.org/https://doi.org/10.1016/S0168-9002(02)01274-3} {\bibfield
  {journal} {\bibinfo  {journal} {Nucl. Instr. Meth. A}\ }\textbf {\bibinfo
  {volume} {491}},\ \bibinfo {pages} {460 } (\bibinfo {year}
  {2002})}\BibitemShut {NoStop}%
\bibitem [{\citenamefont {Luo}\ \emph {et~al.}(2014)\citenamefont {Luo} \emph
  {et~al.}}]{irt1}%
  \BibitemOpen
  \bibfield  {author} {\bibinfo {author} {\bibfnamefont {X.~L.}\ \bibnamefont
  {Luo}} \emph {et~al.},\ }\href
  {https://doi.org/https://doi.org/10.1016/j.nima.2014.08.023} {\bibfield
  {journal} {\bibinfo  {journal} {Nucl. Instr. Meth. A}\ }\textbf {\bibinfo
  {volume} {767}},\ \bibinfo {pages} {83 } (\bibinfo {year}
  {2014})}\BibitemShut {NoStop}%
\bibitem [{\citenamefont {Ronchi}\ \emph {et~al.}(2009)\citenamefont {Ronchi},
  \citenamefont {Soderstrom}, \citenamefont {Nyberg}, \citenamefont {Sunden},
  \citenamefont {Conroy}, \citenamefont {Ericsson}, \citenamefont {Hellesen},
  \citenamefont {Johnson},\ and\ \citenamefont {Weiszflog}}]{irt2}%
  \BibitemOpen
  \bibfield  {author} {\bibinfo {author} {\bibfnamefont {E.}~\bibnamefont
  {Ronchi}}, \bibinfo {author} {\bibfnamefont {P.-A.}\ \bibnamefont
  {Soderstrom}}, \bibinfo {author} {\bibfnamefont {J.}~\bibnamefont {Nyberg}},
  \bibinfo {author} {\bibfnamefont {E.~A.}\ \bibnamefont {Sunden}}, \bibinfo
  {author} {\bibfnamefont {S.}~\bibnamefont {Conroy}}, \bibinfo {author}
  {\bibfnamefont {G.}~\bibnamefont {Ericsson}}, \bibinfo {author}
  {\bibfnamefont {C.}~\bibnamefont {Hellesen}}, \bibinfo {author}
  {\bibfnamefont {M.~G.}\ \bibnamefont {Johnson}},\ and\ \bibinfo {author}
  {\bibfnamefont {M.}~\bibnamefont {Weiszflog}},\ }\href
  {https://doi.org/https://doi.org/10.1016/j.nima.2009.08.064} {\bibfield
  {journal} {\bibinfo  {journal} {Nucl. Instr. Meth. A}\ }\textbf {\bibinfo
  {volume} {610}},\ \bibinfo {pages} {534 } (\bibinfo {year}
  {2009})}\BibitemShut {NoStop}%
\bibitem [{\citenamefont {Pozzi}\ \emph {et~al.}(2003)\citenamefont {Pozzi},
  \citenamefont {Padovani},\ and\ \citenamefont {Marseguerra}}]{polimi}%
  \BibitemOpen
  \bibfield  {author} {\bibinfo {author} {\bibfnamefont {S.~A.}\ \bibnamefont
  {Pozzi}}, \bibinfo {author} {\bibfnamefont {E.}~\bibnamefont {Padovani}},\
  and\ \bibinfo {author} {\bibfnamefont {M.}~\bibnamefont {Marseguerra}},\
  }\href {https://doi.org/https://doi.org/10.1016/j.nima.2003.06.012}
  {\bibfield  {journal} {\bibinfo  {journal} {Nucl. Instr. Meth. A}\ }\textbf
  {\bibinfo {volume} {513}},\ \bibinfo {pages} {550 } (\bibinfo {year}
  {2003})}\BibitemShut {NoStop}%
\bibitem [{\citenamefont {Awe}\ \emph {et~al.}(2018)\citenamefont {Awe},
  \citenamefont {Barbeau}, \citenamefont {Collar}, \citenamefont {Hedges},\
  and\ \citenamefont {Li}}]{ej301}%
  \BibitemOpen
  \bibfield  {author} {\bibinfo {author} {\bibfnamefont {C.}~\bibnamefont
  {Awe}}, \bibinfo {author} {\bibfnamefont {P.~S.}\ \bibnamefont {Barbeau}},
  \bibinfo {author} {\bibfnamefont {J.~I.}\ \bibnamefont {Collar}}, \bibinfo
  {author} {\bibfnamefont {S.}~\bibnamefont {Hedges}},\ and\ \bibinfo {author}
  {\bibfnamefont {L.}~\bibnamefont {Li}},\ }\href
  {https://doi.org/10.1103/PhysRevC.98.045802} {\bibfield  {journal} {\bibinfo
  {journal} {Phys. Rev. C}\ }\textbf {\bibinfo {volume} {98}},\ \bibinfo
  {pages} {045802} (\bibinfo {year} {2018})}\BibitemShut {NoStop}%
\bibitem [{\citenamefont {Hutcheson}\ \emph {et~al.}(2007)\citenamefont
  {Hutcheson} \emph {et~al.}}]{tunldd}%
  \BibitemOpen
  \bibfield  {author} {\bibinfo {author} {\bibfnamefont {A.}~\bibnamefont
  {Hutcheson}} \emph {et~al.},\ }\href
  {https://doi.org/https://doi.org/10.1016/j.nimb.2007.04.097} {\bibfield
  {journal} {\bibinfo  {journal} {Nucl. Instr. Meth. B}\ }\textbf {\bibinfo
  {volume} {261}},\ \bibinfo {pages} {369 } (\bibinfo {year}
  {2007})}\BibitemShut {NoStop}%
\bibitem [{han()}]{handbook}%
  \BibitemOpen
  \href@noop {} {}\bibinfo {note} {Photomultiplier Tubes: Basics and
  Applications. \\ \url{https://www.hamamatsu.com}}\BibitemShut {NoStop}%
\bibitem [{\citenamefont {Kossakowski}\ \emph {et~al.}(2002)\citenamefont
  {Kossakowski}, \citenamefont {Audemer}, \citenamefont {Dubois}, \citenamefont
  {Fougeron}, \citenamefont {Hermel}, \citenamefont {Sottile},\ and\
  \citenamefont {Vialle}}]{report}%
  \BibitemOpen
  \bibfield  {author} {\bibinfo {author} {\bibfnamefont {R.}~\bibnamefont
  {Kossakowski}}, \bibinfo {author} {\bibfnamefont {J.~C.}\ \bibnamefont
  {Audemer}}, \bibinfo {author} {\bibfnamefont {J.~M.}\ \bibnamefont {Dubois}},
  \bibinfo {author} {\bibfnamefont {D.}~\bibnamefont {Fougeron}}, \bibinfo
  {author} {\bibfnamefont {R.}~\bibnamefont {Hermel}}, \bibinfo {author}
  {\bibfnamefont {R.}~\bibnamefont {Sottile}},\ and\ \bibinfo {author}
  {\bibfnamefont {J.-P.}\ \bibnamefont {Vialle}},\ }\bibfield  {title}
  {\bibinfo {title} {Study of the photomultiplier \protect{R7600-00-M4} for the
  purpose of the electromagnetic calorimeter in the \protect{AMS-02}
  experiment},\ }\href@noop {} {\bibfield  {journal} {\bibinfo  {journal}
  {Report LAPP-EXP--2002-02}\ } (\bibinfo {year} {2002})}\BibitemShut {NoStop}%
\bibitem [{\citenamefont {Keszthelyi-Landori}\ and\ \citenamefont
  {Hrehuss}(1969)}]{decay}%
  \BibitemOpen
  \bibfield  {author} {\bibinfo {author} {\bibfnamefont {S.}~\bibnamefont
  {Keszthelyi-Landori}}\ and\ \bibinfo {author} {\bibfnamefont
  {G.}~\bibnamefont {Hrehuss}},\ }\href
  {https://doi.org/https://doi.org/10.1016/0029-554X(69)90682-X} {\bibfield
  {journal} {\bibinfo  {journal} {Nucl. Instr. Meth.}\ }\textbf {\bibinfo
  {volume} {68}},\ \bibinfo {pages} {9 } (\bibinfo {year} {1969})}\BibitemShut
  {NoStop}%
\bibitem [{pek()}]{pekev}%
  \BibitemOpen
  \href@noop {} {}\bibinfo {note} {\protect{Mention of $^{241}$Am light yield
  in \cite{grayson} erroneously uses PE/keV units, where scintillation photons
  per keV were intended. The correct 12.0 PE/keV yield is obtained via the 40\%
  quantum efficiency of the UBA photocathode at the CsI[Na] wavelength of
  emission. An independent analysis confirms this value (A. Konovalov, private
  communication).}}\BibitemShut {Stop}%
\bibitem [{\citenamefont {Aprile}\ \emph {et~al.}(2018)\citenamefont {Aprile},
  \citenamefont {Anthony}, \citenamefont {Lin}, \citenamefont {Greene},
  \citenamefont {de~Perio}, \citenamefont {Gao}, \citenamefont {Howlett},
  \citenamefont {Plante}, \citenamefont {Zhang},\ and\ \citenamefont
  {Zhu}}]{leff}%
  \BibitemOpen
  \bibfield  {author} {\bibinfo {author} {\bibfnamefont {E.}~\bibnamefont
  {Aprile}}, \bibinfo {author} {\bibfnamefont {M.}~\bibnamefont {Anthony}},
  \bibinfo {author} {\bibfnamefont {Q.}~\bibnamefont {Lin}}, \bibinfo {author}
  {\bibfnamefont {Z.}~\bibnamefont {Greene}}, \bibinfo {author} {\bibfnamefont
  {P.}~\bibnamefont {de~Perio}}, \bibinfo {author} {\bibfnamefont
  {F.}~\bibnamefont {Gao}}, \bibinfo {author} {\bibfnamefont {J.}~\bibnamefont
  {Howlett}}, \bibinfo {author} {\bibfnamefont {G.}~\bibnamefont {Plante}},
  \bibinfo {author} {\bibfnamefont {Y.}~\bibnamefont {Zhang}},\ and\ \bibinfo
  {author} {\bibfnamefont {T.}~\bibnamefont {Zhu}},\ }\href
  {https://doi.org/10.1103/PhysRevD.98.112003} {\bibfield  {journal} {\bibinfo
  {journal} {Phys. Rev. D}\ }\textbf {\bibinfo {volume} {98}},\ \bibinfo
  {pages} {112003} (\bibinfo {year} {2018})}\BibitemShut {NoStop}%
\bibitem [{\citenamefont {{Beck}}\ \emph {et~al.}(2015)\citenamefont {{Beck}},
  \citenamefont {{Payne}}, \citenamefont {{Hunter}}, \citenamefont {{Ahle}},
  \citenamefont {{Cherepy}},\ and\ \citenamefont {{Swanberg}}}]{nonprop1}%
  \BibitemOpen
  \bibfield  {author} {\bibinfo {author} {\bibfnamefont {P.~R.}\ \bibnamefont
  {{Beck}}}, \bibinfo {author} {\bibfnamefont {S.~A.}\ \bibnamefont {{Payne}}},
  \bibinfo {author} {\bibfnamefont {S.}~\bibnamefont {{Hunter}}}, \bibinfo
  {author} {\bibfnamefont {L.}~\bibnamefont {{Ahle}}}, \bibinfo {author}
  {\bibfnamefont {N.~J.}\ \bibnamefont {{Cherepy}}},\ and\ \bibinfo {author}
  {\bibfnamefont {E.~L.}\ \bibnamefont {{Swanberg}}},\ }\href
  {https://doi.org/10.1109/TNS.2015.2414357} {\bibfield  {journal} {\bibinfo
  {journal} {IEEE Trans. Nucl. Sci.}\ }\textbf {\bibinfo {volume} {62}},\
  \bibinfo {pages} {1429} (\bibinfo {year} {2015})}\BibitemShut {NoStop}%
\bibitem [{\citenamefont {Salakhutdinov}\ and\ \citenamefont
  {Efanov}(2015)}]{nonprop2}%
  \BibitemOpen
  \bibfield  {author} {\bibinfo {author} {\bibfnamefont {G.~K.}\ \bibnamefont
  {Salakhutdinov}}\ and\ \bibinfo {author} {\bibfnamefont {D.~V.}\ \bibnamefont
  {Efanov}},\ }\href {https://doi.org/10.1134/S0020441215030100} {\bibfield
  {journal} {\bibinfo  {journal} {Instrum. Exp. Tech.}\ }\textbf {\bibinfo
  {volume} {58}},\ \bibinfo {pages} {345} (\bibinfo {year} {2015})}\BibitemShut
  {NoStop}%
\bibitem [{\citenamefont {{Mengesha}}\ \emph {et~al.}(1998)\citenamefont
  {{Mengesha}}, \citenamefont {{Taulbee}}, \citenamefont {{Rooney}},\ and\
  \citenamefont {{Valentine}}}]{nonprop3}%
  \BibitemOpen
  \bibfield  {author} {\bibinfo {author} {\bibfnamefont {W.}~\bibnamefont
  {{Mengesha}}}, \bibinfo {author} {\bibfnamefont {T.~D.}\ \bibnamefont
  {{Taulbee}}}, \bibinfo {author} {\bibfnamefont {B.~D.}\ \bibnamefont
  {{Rooney}}},\ and\ \bibinfo {author} {\bibfnamefont {J.~D.}\ \bibnamefont
  {{Valentine}}},\ }\href {https://doi.org/10.1109/23.682426} {\bibfield
  {journal} {\bibinfo  {journal} {IEEE Trans. Nucl. Sci.}\ }\textbf {\bibinfo
  {volume} {45}},\ \bibinfo {pages} {456} (\bibinfo {year} {1998})}\BibitemShut
  {NoStop}%
\bibitem [{\citenamefont {Tretyak}(2010)}]{tretyak}%
  \BibitemOpen
  \bibfield  {author} {\bibinfo {author} {\bibfnamefont {V.}~\bibnamefont
  {Tretyak}},\ }\href
  {https://doi.org/https://doi.org/10.1016/j.astropartphys.2009.11.002}
  {\bibfield  {journal} {\bibinfo  {journal} {Astropart. Phys.}\ }\textbf
  {\bibinfo {volume} {33}},\ \bibinfo {pages} {40 } (\bibinfo {year}
  {2010})}\BibitemShut {NoStop}%
\bibitem [{\citenamefont {Birks}(1951)}]{birks}%
  \BibitemOpen
  \bibfield  {author} {\bibinfo {author} {\bibfnamefont {J.~B.}\ \bibnamefont
  {Birks}},\ }\href {https://doi.org/10.1088/0370-1298/64/10/303} {\bibfield
  {journal} {\bibinfo  {journal} {Procs. Phys. Soc. Sect. A}\ }\textbf
  {\bibinfo {volume} {64}},\ \bibinfo {pages} {874} (\bibinfo {year}
  {1951})}\BibitemShut {NoStop}%
\bibitem [{\citenamefont {Ficenec}\ \emph {et~al.}(1987)\citenamefont
  {Ficenec}, \citenamefont {Ahlen}, \citenamefont {Marin}, \citenamefont
  {Musser},\ and\ \citenamefont {Tarl\'e}}]{ahlen1}%
  \BibitemOpen
  \bibfield  {author} {\bibinfo {author} {\bibfnamefont {D.~J.}\ \bibnamefont
  {Ficenec}}, \bibinfo {author} {\bibfnamefont {S.~P.}\ \bibnamefont {Ahlen}},
  \bibinfo {author} {\bibfnamefont {A.~A.}\ \bibnamefont {Marin}}, \bibinfo
  {author} {\bibfnamefont {J.~A.}\ \bibnamefont {Musser}},\ and\ \bibinfo
  {author} {\bibfnamefont {G.}~\bibnamefont {Tarl\'e}},\ }\href
  {https://doi.org/10.1103/PhysRevD.36.311} {\bibfield  {journal} {\bibinfo
  {journal} {Phys. Rev. D}\ }\textbf {\bibinfo {volume} {36}},\ \bibinfo
  {pages} {311} (\bibinfo {year} {1987})}\BibitemShut {NoStop}%
\bibitem [{\citenamefont {Ahlen}\ and\ \citenamefont {Tarl\'e}(1983)}]{ahlen2}%
  \BibitemOpen
  \bibfield  {author} {\bibinfo {author} {\bibfnamefont {S.~P.}\ \bibnamefont
  {Ahlen}}\ and\ \bibinfo {author} {\bibfnamefont {G.}~\bibnamefont
  {Tarl\'e}},\ }\href {https://doi.org/10.1103/PhysRevD.27.688} {\bibfield
  {journal} {\bibinfo  {journal} {Phys. Rev. D}\ }\textbf {\bibinfo {volume}
  {27}},\ \bibinfo {pages} {688} (\bibinfo {year} {1983})}\BibitemShut
  {NoStop}%
\bibitem [{\citenamefont {Ziegler}(1999)}]{ziegler}%
  \BibitemOpen
  \bibfield  {author} {\bibinfo {author} {\bibfnamefont {J.~F.}\ \bibnamefont
  {Ziegler}},\ }\href {https://doi.org/10.1063/1.369844} {\bibfield  {journal}
  {\bibinfo  {journal} {J. Appl. Phys.}\ }\textbf {\bibinfo {volume} {85}},\
  \bibinfo {pages} {1249} (\bibinfo {year} {1999})}\BibitemShut {NoStop}%
\bibitem [{sri()}]{srim}%
  \BibitemOpen
  \href@noop {} {}\bibinfo {howpublished}
  {\url{http://www.srim.org/}}\BibitemShut {NoStop}%
\bibitem [{\citenamefont {Foreman-Mackey}\ \emph {et~al.}(2013)\citenamefont
  {Foreman-Mackey}, \citenamefont {Hogg}, \citenamefont {Lang},\ and\
  \citenamefont {Goodman}}]{mcmc1}%
  \BibitemOpen
  \bibfield  {author} {\bibinfo {author} {\bibfnamefont {D.}~\bibnamefont
  {Foreman-Mackey}}, \bibinfo {author} {\bibfnamefont {D.~W.}\ \bibnamefont
  {Hogg}}, \bibinfo {author} {\bibfnamefont {D.}~\bibnamefont {Lang}},\ and\
  \bibinfo {author} {\bibfnamefont {J.}~\bibnamefont {Goodman}},\ }\href
  {https://doi.org/10.1086/670067} {\bibfield  {journal} {\bibinfo  {journal}
  {Publ. Astron. Soc. Pac.}\ }\textbf {\bibinfo {volume} {125}},\ \bibinfo
  {pages} {306} (\bibinfo {year} {2013})}\BibitemShut {NoStop}%
\bibitem [{mcm()}]{mcmc2}%
  \BibitemOpen
  \href@noop {} {}\bibinfo {howpublished}
  {\url{https://emcee.readthedocs.io/en/v2.2.1/}}\BibitemShut {NoStop}%
\bibitem [{\citenamefont {Ong}\ \emph {et~al.}(1979)\citenamefont {Ong},
  \citenamefont {Song}, \citenamefont {Monnier},\ and\ \citenamefont
  {Stoneham}}]{bandgap1}%
  \BibitemOpen
  \bibfield  {author} {\bibinfo {author} {\bibfnamefont {C.~K.}\ \bibnamefont
  {Ong}}, \bibinfo {author} {\bibfnamefont {K.~S.}\ \bibnamefont {Song}},
  \bibinfo {author} {\bibfnamefont {R.}~\bibnamefont {Monnier}},\ and\ \bibinfo
  {author} {\bibfnamefont {A.~M.}\ \bibnamefont {Stoneham}},\ }\href
  {https://doi.org/10.1088/0022-3719/12/21/028} {\bibfield  {journal} {\bibinfo
   {journal} {J. Phys. C: Sol. State Phys.}\ }\textbf {\bibinfo {volume}
  {12}},\ \bibinfo {pages} {4641} (\bibinfo {year} {1979})}\BibitemShut
  {NoStop}%
\bibitem [{\citenamefont {Gridin}(2014)}]{bandgap2}%
  \BibitemOpen
  \bibfield  {author} {\bibinfo {author} {\bibfnamefont {S.}~\bibnamefont
  {Gridin}},\ }\href {https://doi.org/10.13140/2.1.4568.4000} {Ph.D. thesis},\
  \bibinfo  {school} {Universit\'e Claude Bernard} (\bibinfo {year}
  {2014})\BibitemShut {NoStop}%
\bibitem [{\citenamefont {Sierra}\ \emph {et~al.}(2019)\citenamefont {Sierra},
  \citenamefont {Liao},\ and\ \citenamefont {Marfatia}}]{danny}%
  \BibitemOpen
  \bibfield  {author} {\bibinfo {author} {\bibfnamefont {D.~A.}\ \bibnamefont
  {Sierra}}, \bibinfo {author} {\bibfnamefont {J.}~\bibnamefont {Liao}},\ and\
  \bibinfo {author} {\bibfnamefont {D.}~\bibnamefont {Marfatia}},\ }\href
  {https://doi.org/10.1007/JHEP06(2019)141} {\bibfield  {journal} {\bibinfo
  {journal} {JHEP}\ }\textbf {\bibinfo {volume} {2019}}\bibinfo  {number} {
  (6)},\ \bibinfo {pages} {141}}\BibitemShut {NoStop}%
\bibitem [{\citenamefont {Coloma}\ \emph {et~al.}(2017)\citenamefont {Coloma},
  \citenamefont {Gonzalez-Garcia}, \citenamefont {Maltoni},\ and\ \citenamefont
  {Schwetz}}]{lma}%
  \BibitemOpen
\bibfield  {number} {  }\bibfield  {author} {\bibinfo {author} {\bibfnamefont
  {P.}~\bibnamefont {Coloma}}, \bibinfo {author} {\bibfnamefont {M.~C.}\
  \bibnamefont {Gonzalez-Garcia}}, \bibinfo {author} {\bibfnamefont
  {M.}~\bibnamefont {Maltoni}},\ and\ \bibinfo {author} {\bibfnamefont
  {T.}~\bibnamefont {Schwetz}},\ }\href
  {https://doi.org/10.1103/PhysRevD.96.115007} {\bibfield  {journal} {\bibinfo
  {journal} {Phys. Rev. D}\ }\textbf {\bibinfo {volume} {96}},\ \bibinfo
  {pages} {115007} (\bibinfo {year} {2017})}\BibitemShut {NoStop}%
\bibitem [{\citenamefont {Denton}\ \emph {et~al.}(2018)\citenamefont {Denton},
  \citenamefont {Farzan},\ and\ \citenamefont {Shoemaker}}]{ian}%
  \BibitemOpen
  \bibfield  {author} {\bibinfo {author} {\bibfnamefont {P.~B.}\ \bibnamefont
  {Denton}}, \bibinfo {author} {\bibfnamefont {Y.}~\bibnamefont {Farzan}},\
  and\ \bibinfo {author} {\bibfnamefont {I.~M.}\ \bibnamefont {Shoemaker}},\
  }\href@noop {} {\  (\bibinfo {year} {2018})},\ \Eprint
  {https://arxiv.org/abs/1804.03660} {arXiv:1804.03660} \BibitemShut {NoStop}%
\bibitem [{\citenamefont {Huang}\ and\ \citenamefont {Chen}(2019)}]{huang1}%
  \BibitemOpen
  \bibfield  {author} {\bibinfo {author} {\bibfnamefont {X.-R.}\ \bibnamefont
  {Huang}}\ and\ \bibinfo {author} {\bibfnamefont {L.-W.}\ \bibnamefont
  {Chen}},\ }\href@noop {} {\  (\bibinfo {year} {2019})},\ \Eprint
  {https://arxiv.org/abs/1902.07625} {arXiv:1902.07625} \BibitemShut {NoStop}%
\bibitem [{hua()}]{huang2}%
  \BibitemOpen
  \href@noop {} {}\bibinfo {note} {X.-R. Huang and L.-W. Chen, private
  communication.}\BibitemShut {Stop}%
\bibitem [{\citenamefont {Dutta}\ \emph {et~al.}(2019)\citenamefont {Dutta},
  \citenamefont {Kim}, \citenamefont {Park}, \citenamefont {Shin},\ and\
  \citenamefont {Strigari}}]{strigari}%
  \BibitemOpen
  \bibfield  {author} {\bibinfo {author} {\bibfnamefont {B.}~\bibnamefont
  {Dutta}}, \bibinfo {author} {\bibfnamefont {D.}~\bibnamefont {Kim}}, \bibinfo
  {author} {\bibfnamefont {J.-C.}\ \bibnamefont {Park}}, \bibinfo {author}
  {\bibfnamefont {S.}~\bibnamefont {Shin}},\ and\ \bibinfo {author}
  {\bibfnamefont {L.~E.}\ \bibnamefont {Strigari}},\ }\href@noop {} {\
  (\bibinfo {year} {2019})},\ \Eprint {https://arxiv.org/abs/1906.10745}
  {arXiv:1906.10745} \BibitemShut {NoStop}%
\bibitem [{\citenamefont {Bednyakov}\ and\ \citenamefont
  {Naumov}(2018)}]{incoherent}%
  \BibitemOpen
  \bibfield  {author} {\bibinfo {author} {\bibfnamefont {V.~A.}\ \bibnamefont
  {Bednyakov}}\ and\ \bibinfo {author} {\bibfnamefont {D.~V.}\ \bibnamefont
  {Naumov}},\ }\href {https://doi.org/10.1103/PhysRevD.98.053004} {\bibfield
  {journal} {\bibinfo  {journal} {Phys. Rev. D}\ }\textbf {\bibinfo {volume}
  {98}},\ \bibinfo {pages} {053004} (\bibinfo {year} {2018})}\BibitemShut
  {NoStop}%
\bibitem [{\citenamefont {Barbeau}\ \emph
  {et~al.}(2007{\natexlab{a}})\citenamefont {Barbeau}, \citenamefont {Collar},\
  and\ \citenamefont {Tench}}]{ppc}%
  \BibitemOpen
  \bibfield  {author} {\bibinfo {author} {\bibfnamefont {P.~S.}\ \bibnamefont
  {Barbeau}}, \bibinfo {author} {\bibfnamefont {J.~I.}\ \bibnamefont
  {Collar}},\ and\ \bibinfo {author} {\bibfnamefont {O.}~\bibnamefont
  {Tench}},\ }\href {https://doi.org/10.1088/1475-7516/2007/09/009} {\bibfield
  {journal} {\bibinfo  {journal} {JCAP}\ }\textbf {\bibinfo {volume}
  {2007}}\bibinfo  {number} { (09)},\ \bibinfo {pages} {009}}\BibitemShut
  {NoStop}%
\bibitem [{\citenamefont {Collar}(2013{\natexlab{b}})}]{ybeprl}%
  \BibitemOpen
\bibfield  {number} {  }\bibfield  {author} {\bibinfo {author} {\bibfnamefont
  {J.~I.}\ \bibnamefont {Collar}},\ }\href
  {https://doi.org/10.1103/PhysRevLett.110.211101} {\bibfield  {journal}
  {\bibinfo  {journal} {Phys. Rev. Lett.}\ }\textbf {\bibinfo {volume} {110}},\
  \bibinfo {pages} {211101} (\bibinfo {year} {2013}{\natexlab{b}})}\BibitemShut
  {NoStop}%
\bibitem [{\citenamefont {Scholz}\ \emph {et~al.}(2016)\citenamefont {Scholz},
  \citenamefont {Chavarria}, \citenamefont {Collar}, \citenamefont
  {Privitera},\ and\ \citenamefont {Robinson}}]{bjornqf}%
  \BibitemOpen
  \bibfield  {author} {\bibinfo {author} {\bibfnamefont {B.~J.}\ \bibnamefont
  {Scholz}}, \bibinfo {author} {\bibfnamefont {A.~E.}\ \bibnamefont
  {Chavarria}}, \bibinfo {author} {\bibfnamefont {J.~I.}\ \bibnamefont
  {Collar}}, \bibinfo {author} {\bibfnamefont {P.}~\bibnamefont {Privitera}},\
  and\ \bibinfo {author} {\bibfnamefont {A.~E.}\ \bibnamefont {Robinson}},\
  }\href {https://doi.org/10.1103/PhysRevD.94.122003} {\bibfield  {journal}
  {\bibinfo  {journal} {Phys. Rev. D}\ }\textbf {\bibinfo {volume} {94}},\
  \bibinfo {pages} {122003} (\bibinfo {year} {2016})}\BibitemShut {NoStop}%
\bibitem [{\citenamefont {Chavarria}\ \emph {et~al.}(2016)\citenamefont
  {Chavarria} \emph {et~al.}}]{alvaro}%
  \BibitemOpen
  \bibfield  {author} {\bibinfo {author} {\bibfnamefont {A.~E.}\ \bibnamefont
  {Chavarria}} \emph {et~al.},\ }\href
  {https://doi.org/10.1103/PhysRevD.94.082007} {\bibfield  {journal} {\bibinfo
  {journal} {Phys. Rev. D}\ }\textbf {\bibinfo {volume} {94}},\ \bibinfo
  {pages} {082007} (\bibinfo {year} {2016})}\BibitemShut {NoStop}%
\bibitem [{\citenamefont {Jones}\ and\ \citenamefont {Kraner}(1975)}]{jones}%
  \BibitemOpen
  \bibfield  {author} {\bibinfo {author} {\bibfnamefont {K.~W.}\ \bibnamefont
  {Jones}}\ and\ \bibinfo {author} {\bibfnamefont {H.~W.}\ \bibnamefont
  {Kraner}},\ }\href {https://doi.org/10.1103/PhysRevA.11.1347} {\bibfield
  {journal} {\bibinfo  {journal} {Phys. Rev. A}\ }\textbf {\bibinfo {volume}
  {11}},\ \bibinfo {pages} {1347} (\bibinfo {year} {1975})}\BibitemShut
  {NoStop}%
\bibitem [{\citenamefont {Turkoglu}\ \emph {et~al.}(2012)\citenamefont
  {Turkoglu}, \citenamefont {Burke}, \citenamefont {Lewandowski},\ and\
  \citenamefont {Cao}}]{ohio}%
  \BibitemOpen
  \bibfield  {author} {\bibinfo {author} {\bibfnamefont {D.}~\bibnamefont
  {Turkoglu}}, \bibinfo {author} {\bibfnamefont {J.}~\bibnamefont {Burke}},
  \bibinfo {author} {\bibfnamefont {R.}~\bibnamefont {Lewandowski}},\ and\
  \bibinfo {author} {\bibfnamefont {L.~R.}\ \bibnamefont {Cao}},\ }\href
  {https://doi.org/10.1007/s10967-011-1289-2} {\bibfield  {journal} {\bibinfo
  {journal} {J. Radioanal. Nucl. Chem.}\ }\textbf {\bibinfo {volume} {291}},\
  \bibinfo {pages} {321} (\bibinfo {year} {2012})}\BibitemShut {NoStop}%
\bibitem [{\citenamefont {Barbeau}\ \emph
  {et~al.}(2007{\natexlab{b}})\citenamefont {Barbeau}, \citenamefont {Collar},\
  and\ \citenamefont {Whaley}}]{ksu}%
  \BibitemOpen
  \bibfield  {author} {\bibinfo {author} {\bibfnamefont {P.~S.}\ \bibnamefont
  {Barbeau}}, \bibinfo {author} {\bibfnamefont {J.~I.}\ \bibnamefont
  {Collar}},\ and\ \bibinfo {author} {\bibfnamefont {P.~M.}\ \bibnamefont
  {Whaley}},\ }\href
  {https://doi.org/https://doi.org/10.1016/j.nima.2007.01.169} {\bibfield
  {journal} {\bibinfo  {journal} {Nucl. Instr. Meth. A}\ }\textbf {\bibinfo
  {volume} {574}},\ \bibinfo {pages} {385 } (\bibinfo {year}
  {2007}{\natexlab{b}})}\BibitemShut {NoStop}%
\bibitem [{\citenamefont {Jiang}\ \emph {et~al.}(2018)\citenamefont {Jiang}
  \emph {et~al.}}]{cdex}%
  \BibitemOpen
  \bibfield  {author} {\bibinfo {author} {\bibfnamefont {H.}~\bibnamefont
  {Jiang}} \emph {et~al.} (\bibinfo {collaboration} {CDEX Collaboration}),\
  }\href {https://doi.org/10.1103/PhysRevLett.120.241301} {\bibfield  {journal}
  {\bibinfo  {journal} {Phys. Rev. Lett.}\ }\textbf {\bibinfo {volume} {120}},\
  \bibinfo {pages} {241301} (\bibinfo {year} {2018})}\BibitemShut {NoStop}%
\bibitem [{\citenamefont {Auerbach}\ \emph {et~al.}(2001)\citenamefont
  {Auerbach} \emph {et~al.}}]{lsnd}%
  \BibitemOpen
  \bibfield  {author} {\bibinfo {author} {\bibfnamefont {L.~B.}\ \bibnamefont
  {Auerbach}} \emph {et~al.} (\bibinfo {collaboration} {LSND Collaboration}),\
  }\href {https://doi.org/10.1103/PhysRevD.63.112001} {\bibfield  {journal}
  {\bibinfo  {journal} {Phys. Rev. D}\ }\textbf {\bibinfo {volume} {63}},\
  \bibinfo {pages} {112001} (\bibinfo {year} {2001})}\BibitemShut {NoStop}%
\end{thebibliography}%

\end{document}